\documentclass[twocolumn,linenumbers,trackchanges]{aastex62}

\usepackage{array}
\usepackage{CJKutf8}

\newcommand{\cntext}[1]{\begin{CJK}{UTF8}{gbsn}#1\end{CJK}\kern-1ex}

\graphicspath{{./}{figures/}}
\turnoffeditone
\nolinenumbers

\received{2022 July 10}
\revised{2022 September 21}
\accepted{2022 October 10}
\published{2022 November 29}
\submitjournal{ApJ}

\begin{document}

\title{Multiple Regions of Nonthermal Quasiperiodic Pulsations during the Impulsive Phase of a Solar Flare}

\correspondingauthor{Yingjie Luo}
\email{yl863@njit.edu}

\author[0000-0002-5431-545X]{Yingjie Luo (\cntext{骆英杰})}
\affiliation{Center for Solar-Terrestrial Research, New Jersey Institute of Technology, 323 Martin Luther King Jr Blvd., Newark, NJ 07102-1982, USA}

\author[0000-0002-0660-3350]{Bin Chen (\cntext{陈彬})}
\affiliation{Center for Solar-Terrestrial Research, New Jersey Institute of Technology, 323 Martin Luther King Jr Blvd., Newark, NJ 07102-1982, USA}

\author[0000-0003-2872-2614]{Sijie Yu (\cntext{余思捷})}
\affiliation{Center for Solar-Terrestrial Research, New Jersey Institute of Technology, 323 Martin Luther King Jr Blvd., Newark, NJ 07102-1982, USA}

\author[0000-0003-1438-9099]{Marina Battaglia}
\affiliation{University of Applied Sciences and Arts Northwestern Switzerland, 5210 Windisch, Switzerland}

\author[0000-0003-0485-7098]{Rohit Sharma}
\affiliation{University of Applied Sciences and Arts Northwestern Switzerland, 5210 Windisch, Switzerland}

\begin{abstract}

Flare-associated quasiperiodic pulsations (QPPs) in radio and X-ray wavelengths, particularly those related to nonthermal electrons, contain important information about the energy release and transport processes during flares. However, the paucity of spatially resolved observations of such QPPs with a fast time cadence has been an obstacle for us to further understand their physical nature. Here, we report observations of such a QPP event that occurred during the impulsive phase of a C1.8-class eruptive solar flare using radio imaging spectroscopy data from the Karl G. Jansky Very Large Array (VLA) and complementary X-ray imaging and spectroscopy data. The radio QPPs, observed by the VLA in the 1--2 GHz with a subsecond cadence, are shown as three spatially distinct sources with different physical characteristics. Two radio sources are located near the conjugate footpoints of the erupting magnetic flux rope with opposite senses of polarization. One of the sources displays a QPP behavior with a $\sim$5-s period. The third radio source, located at the top of the postflare arcade, coincides with the location of an X-ray source and shares a similar period of $\sim$25--45 s. We show that the two oppositely polarized radio sources are likely due to coherent electron cyclotron maser (ECM) emission. On the other hand, the looptop QPP source, observed in both radio and X-rays, is consistent with incoherent gyrosynchrotron and bremsstrahlung emission, respectively. We conclude that the concurrent, but spatially distinct QPP sources must involve multiple mechanisms which operate in different magnetic loop systems and at different periods.

\nolinenumbers
\end{abstract}


\section{Introduction} \label{sec:intro}

Quasiperiodic pulsations (QPPs) are ubiquitous phenomena in solar flares. They manifest as a time sequence of pulses with a quasiperiodic signature. Since the first report of flare-related QPP event by \citet{1969ApJ...155L.117P}, QPPs have been observed in different wavelengths and flare phases with a variety of properties, including periodicity, duration, and bandwidth (see \citealt{2009SSRv..149..119N,2016SoPh..291.3143V, 2018SSRv..214...45M,2020STP.....6a...3K,2021SSRv..217...66Z} for recent reviews). The periods of QPPs range from seconds to several minutes \citep{2000A&A...360..715K,2014ApJ...791...44H,2016ApJ...827L..30H,2019ApJ...875...33H,2020ApJ...895...50H,2016SoPh..291.3427K,2019ApJ...878...78C,2021ApJ...910..123C}. Reports of QPPs with shorter, subsecond-scale periods are relatively rare (see, e.g., examples in \citealt{2002A&A...385..671F,2012ApJ...749...28T,2013A&A...555A..55Z,2013ApJ...777..159Y}), which may be attributed to the rarity of instruments equipped with subsecond time resolution. 

It has been suggested that QPPs bare important information on the flare energy release, transport, and modulation processes. However, the exact nature of QPPs remains under continued debate. Over fifteen different QPP models in the solar flare picture have been proposed. In a recent review by \citet{2020STP.....6a...3K}, they summarized the different QPP models into three categories. The first category involves modulation of the emissions by magnetohydrodynamics (MHD) oscillations. The second category includes models which suggest that the flare energy release (or its efficiency) can be modulated quasiperiodically by MHD-type oscillations internal or external to the energy release site. The third category attributes the QPPs to emissions due to spontaneous quasiperiodic energy release processes.

While it has become increasingly clear that multiple mechanisms must be at work to interpret the QPPs observed at different wavelengths with a variety of properties, further insights into their origin rely on the availability of spatially, temporally, and spectrally resolved data at multiple wavelengths. \edit1{\citet{2016ApJ...823L..16T} studied a QPP event at EUV wavelengths with a period of 19--27 s. Using spatially and spectrally resolved spectrogram data obtained by the Interface Region Imaging Spectrograph \citep[IRIS;][]{2014SoPh..289.2733D}, they found that the periodic variations occur in both the intensity and the Doppler shift of the \ion{Fe}{21} 1354 \AA\ line. They also noted a phase lag of $\pi/2$ between the intensity and Doppler shift oscillation, which was interpreted as QPPs associated with standing sausage mode MHD waves.} 

Radio observations provide an excellent means for studying QPPs that are intimately related to the flare energy release processes, thanks to their sensitivity to both flare-accelerated nonthermal electrons and heated plasma \citep{1998ARA&A..36..131B}. \edit1{In particular, ground-based radio observations can be made with a very high time cadence, sometimes down to subsecond time scales, offering access to QPPs with very short periods \citep{2010ApJ...723...25T}. In addition, the rich variety of radio emission mechanisms provides the means for diagnosing the quasiperiodic variations of source parameters key to flare energy release, including the magnetic field and nonthermal electron distribution, provided that the responsible emission mechanism can be identified.} 
However, owing to the paucity of simultaneous imaging and spectroscopy capabilities, the vast majority of previously reported radio QPPs have relied on either total-power dynamic spectral data \citep{2012ApJ...748..140M,2013A&A...550A...1K,2016A&A...594A..96G,2018ApJ...855L..29K,2018ApJ...861...33K,2021ApJ...910..123C} or imaging data at a few discrete frequencies \citep{2008ApJ...684.1433F,2009A&A...505..791S,2016ApJ...822....7K,2019ApJ...886L..25Y,2019NatCo..10.2276C}. 

Recently commissioned instruments, including the Karl G. Jansky Very Large Array \citep[VLA;][]{2011ApJ...739L...1P}, Expanded Owens Valley Solar Array \citep[EOVSA;][]{2018ApJ...863...83G}, Mingantu Spectral Radioheliograph \citep[MUSER;][]{2016IAUS..320..427Y}, Murchison Widefield Array \citep[MWA;][]{2013PASA...30....7T}, and Low-Frequency Array \citep[LOFAR;][]{2013A&A...556A...2V}, have enabled studies of solar radio QPPs with true dynamic imaging spectroscopy. For example, at metric wavelengths, using spatially resolved dynamic spectroscopy from MWA, \citet{2019ApJ...883...45M} identified a $\sim$30 s QPP source associated with a microflare. The periodicity of the oscillation is consistent with the Alfv\'{e}n timescale in the typical coronal magnetic environment. The repetitive behavior of the QPP is attributed to a series of reconnection events. Shorter, second-scale QPPs associated with metric type-III-like bursts were also observed during the same event. \citet{2019ApJ...875...98M} suggested that they were likely due to periodic particle injections from the lower corona. However, the studies at long wavelengths are limited by the rather poor angular resolution and the strong propagation effects of the long-wavelength radio waves traversing the corona, rendering the interpretation rather difficult \citep[e.g.][etc.]{1971A&A....10..362S,2019ApJ...884..122K,2020ApJ...903..126S,2022ApJ...932...17Z}. At higher frequencies (decimetric/microwave wavelengths), using imaging spectroscopy data from MUSER, \citet{2019ApJ...878...78C} investigated a QPP event recorded during the impulsive phase of an M8.7-class solar flare in 1.2--2 GHz. The radio QPP event, which had a $\sim$2 minutes period, was located near the null point of the flare with a circular ribbon geometry. It was interpreted as plasma radiation modulated by the flare reconnection. With microwave imaging spectroscopy from EOVSA, \citet{2020ApJ...900...17Y} studied the quasiperiodic radio bursts of a $\sim$300 s period during the post-impulsive phase of the well-studied SOL2017-09-10 X8.2 flare. The recurring thermal looptop emission and nonthermal loopleg emission are closely correlated with the repetitive plasma downflows observed in the large-scale current sheet. They suggested that the quasiperiodic microwave bursts were associated with electron acceleration modulated by the arrival of the plasma downflows at the looptop region.

In this study, by utilizing the radio imaging spectroscopy technique provided by the Karl G. Jansky Very Large Array (VLA), we report that the radio QPPs observed during the impulsive phase of a solar flare are located at spatially distinct source regions with completely different spectral-temporal properties. After a brief overview of the flare event in Section \ref{sec: overview}, we will present in Section \ref{sec: radio} detailed radio observations of the spatially, temporally, and spectrally resolved radio sources by utilizing dynamic imaging spectroscopy. In Section \ref{sec: x-ray}, we will present X-ray observations and the spectral analysis. In Section \ref{sec: analysis}, we place the radio sources into the flare context and discuss the respective emission mechanism for each of the observed radio sources. Finally, in Section \ref{sec: conclusion}, we discuss the possible modulation mechanisms of each QPPs and summarize the results in the context of the flare event.

\section{Observations} \label{sec: observations}

\subsection{Event Overview} \label{sec: overview}

The event under study is a GOES C1.8-class flare that occurred in NOAA active region (AR) 12501 on 2016 February 18. AR 12501 is a long-lasting active region that has survived at least two complete solar rotations. The leading negative sunspot and the surrounding magnetic structure form a null-point topology with a closed fan--dome-type structure. \citet{2019ApJ...874L..33M} reported coronal rains outlining such a null-point topology when the AR was located at the east limb on 2016 March 12 (one rotation later), and the following study \citep{2021ApJ...914L...8M} discussed a variety of successful and failed filament eruptions hosted in the same AR.

\begin{figure*}[!ht]
\plotone{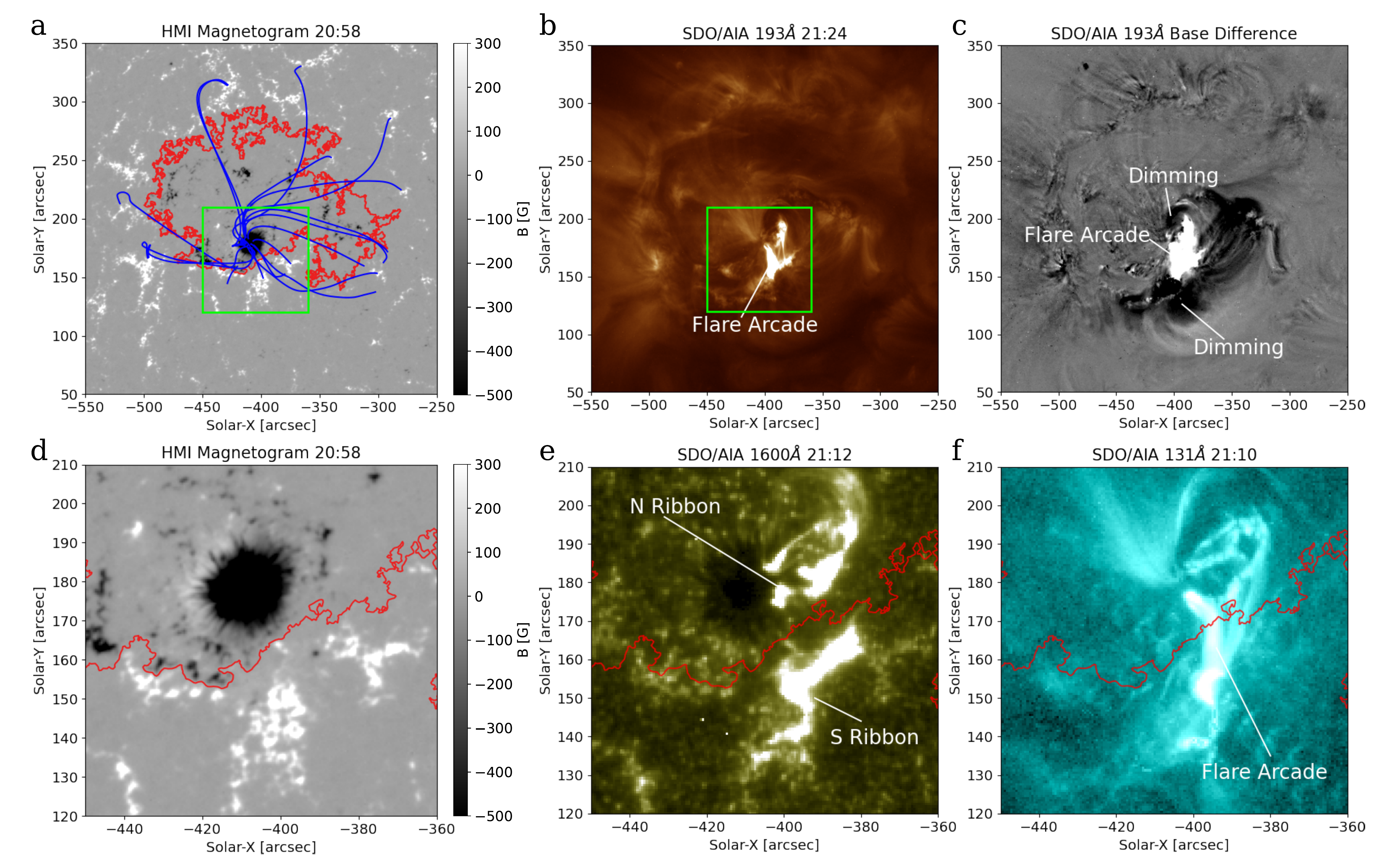}
\caption{Overview of the SOL2016-02-18 C1.8 eruptive solar flare event. (a) SDO/HMI line-of-sight magnetogram of the active region (grayscale background). The polarity inversion line is shown in red. Blue curves show NLFFF extrapolated magnetic field lines, which feature a fan geometry rooted at the central negative sunspot. (b) SDO/AIA 193 \AA\ image of the active region at 21:24 UT. (c) Base-difference SDO/AIA 193 \AA\ image (21:24 UT vs. 21:00 UT) showing the twin coronal dimmings near the conjugate footpoints of the eruption. (d)--(f) Detailed view of the core flare site showing, respectively, the SDO/HMI line-of-sight magnetogram, double flare ribbons as seen in SDO/AIA 1600 \AA, and bright flare arcade observed in SDO/AIA 131 \AA. The field of view is indicated as a green box in (a) and (b). An animation of the eruptive flare as seen in SDO/AIA 193 \AA\ is attached. 
\label{fig: overview}}
\end{figure*}

The site of the flare under study (SOL2016-02-18T21:13) is located at 10$^{\circ}$ north and 24$^{\circ}$ east (in heliographic longitude and latitude) on the solar disk. The photospheric magnetogram, obtained by the Helioseismic and Magnetic Imager \citep[HMI;][]{2012SoPh..275..229S} aboard the Solar Dynamics Observatory \citep[SDO;][]{2012SoPh..275....3P}, shows a negative sunspot surrounded by positive magnetic polarities (Figure \ref{fig: overview}(a)). Nonlinear force-free field (NLFFF) extrapolations based on the SDO/HMI vector magnetogram at 20:58 UT using the method of \citet{2004SoPh..219...87W} show that the magnetic topology features a fan--spine-type geometry, which includes a spine rooted at the central negative sunspot, as well as a fan-shaped dome connecting to the satellite magnetic patches with positive polarity (Figure \ref{fig: overview} (a)). Such a magnetic field configuration is consistent with findings in the previous studies of the same AR \citep{2019ApJ...874L..33M,2021ApJ...914L...8M}. During the course of the flare event, no associated coronal mass ejection (CME) is recorded by the SOHO/LASCO white light coronagraph \citep{2009EM&P..104..295G}. However, two localized EUV coronal dimmings can be observed at both sides of the bright flare arcade as shown by SDO's Atmospheric Imaging Assembly (AIA; \citealt{2012SoPh..275...17L}) 193 \AA\ base-difference images (an example at 21:24 UT is shown in Figure \ref{fig: overview}(c)), which is characteristic of two-ribbon eruptive flares \citep[e.g.,][]{1997ApJ...491L..55S}. These twin dimmings have been interpreted as the signature of the footpoints of an erupting magnetic flux rope where a significant mass-loss is expected through upward expanding plasma flows \citep[e.g.,][]{2001ApJ...561L.215H,2012ApJ...748..106T}---a scenario also supported by recent magnetohydrodynamics (MHD) modeling results \citep{2012ApJ...750..134D,2022ApJ...928..154J}. In our event, one dimming region is located near the edge of the dominant negative sunspot and the other near the nearby positive magnetic polarity. Additionally, two bright flare ribbons also appear in the SDO/AIA 1600 \AA\ images (Figure \ref{fig: overview}(e)). They coincide with the location of the conjugate footpoints of the flare arcade, shown in Figure \ref{fig: overview}(f). The flare arcade first appeared in the 131 \AA\ image during the impulsive phase of the flare ($\sim$21:10 UT). Later on, it cooled down and appeared in \edit1{the other AIA bands that are sensitive to lower coronal temperatures (see Figure \ref{fig: overview}(b) as an example, which shows the AIA 193 \AA\ image at 21:24 UT, or 14 minutes later).} 

\begin{figure*}[!ht]
\plotone{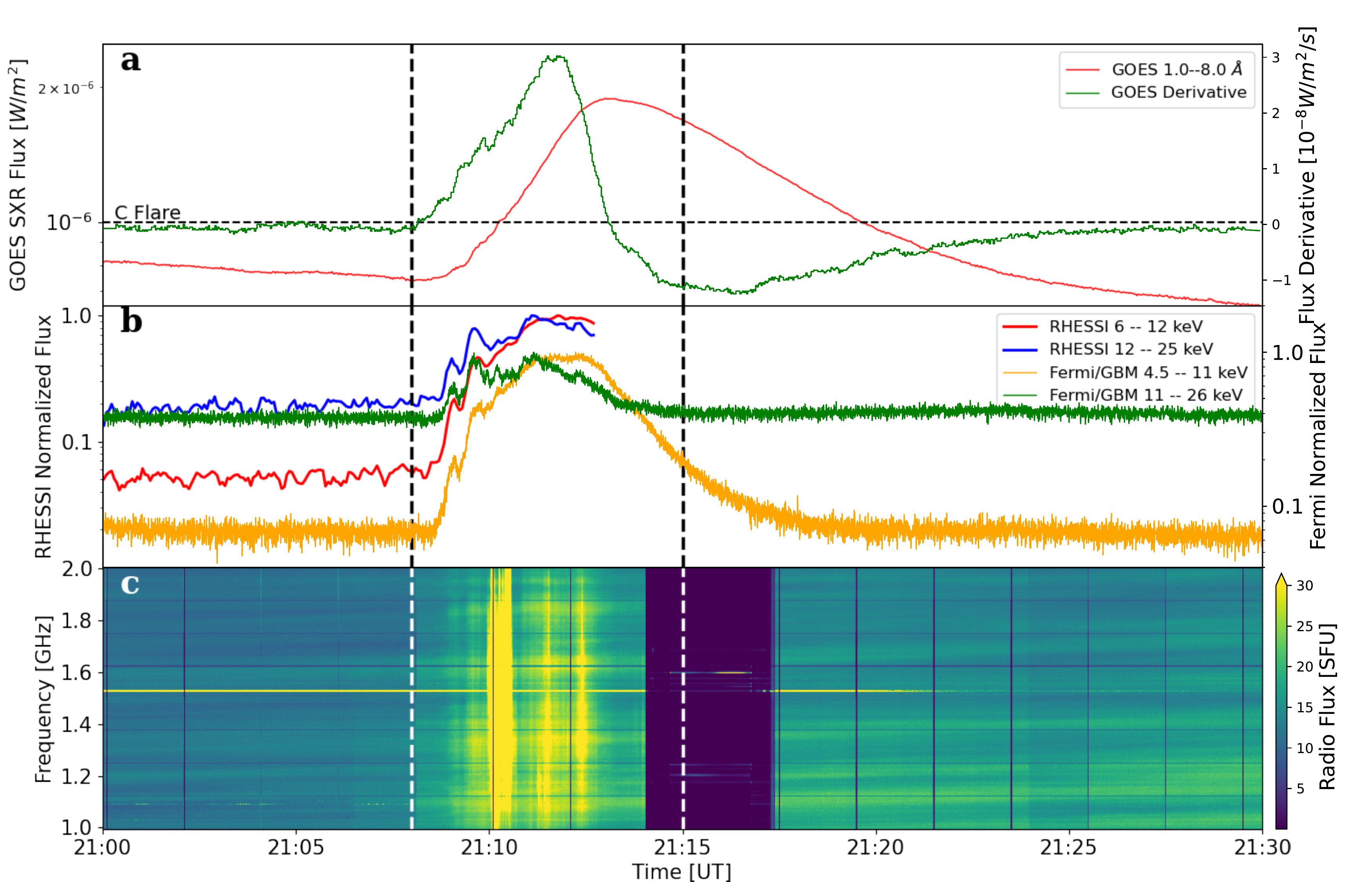}
\caption{X-ray, EUV, and radio time history of the flare. (a) Total-power GOES soft X-ray 1.0--8.0 \AA\ flux (red) and its derivative (green), with the impulsive phase marked between vertical dashed lines. (b) RHESSI 6--12 keV (red) and 12--25 keV (blue) as well as Fermi/GBM 4.5--11 keV (orange) and 11--26 keV (green) light curves. (c) VLA cross-power dynamic spectrum in Stokes I obtained from the median of several selected short baselines\edit1{, shown in solar flux unit, or SFU. 1 $\rm{SFU}=10^4\, \rm{Jy}=10^{-19}\, \rm{erg}\,\rm{s}^{-1}\,\rm{cm}^{-2}\,\rm{Hz}^{-1}$.} \label{fig: lc}}
\end{figure*}

The GOES 1--8 \AA\ soft X-Ray (SXR) flux starts to increase at $\sim$21:09 UT and reaches its peak at $\sim$21:13 UT (red curve in Figure \ref{fig: lc}(a)). The duration of the impulsive phase is relatively short, lasting from $\sim$21:08 to $\sim$21:15 UT (demarcated with vertical dashed lines in Figures \ref{fig: lc}(a)--(c)) according to the GOES 1--8 \AA\ derivative shown as the green curve in Figure \ref{fig: lc}(a). The impulsive phase is partly covered by the Reuven Ramaty High Energy Solar Spectroscopic Imager \citep[RHESSI;][]{2002SoPh..210....3L} and completely observed by the Fermi Gamma-ray Burst Monitor \citep[Fermi/GBM;][]{2009ApJ...702..791M}. Both RHESSI and Fermi/GBM detected multiple X-ray bursts with a quasiperiodic pattern (Figure \ref{fig: lc}(b)). The hard X-ray (HXR) light curves from RHESSI and Fermi/GBM coincide with the derivative of GOES soft X-ray (SXR) flux (Figure \ref{fig: lc}(a) green curve), a typical phenomenon in the flare impulsive phase known as the Neupert effect \citep{1968ApJ...153L..59N,2017LRSP...14....2B}.

VLA radio observations covered the period from 18:59:18 UT on 2016 February 18 to 00:02:14 UT the following day. The observations were made in the C configuration with the full array of 27 antennas \edit1{in the 1--2 GHz $L$ band. The primary beam is large enough to cover the entire solar disk, enabling full-disk imaging}. The longest baseline was 3090 m at the time of observing, which corresponded to an angular resolution of $\sim$15$''$ at 1.5 GHz (which scales inversely with frequency). The observations were performed in the 1--2 GHz with eight 128-MHz-wide spectral windows, each of which is further split into 64 2 MHz wide spectral channels. It enables simultaneous imaging at 512 independent spectral channels and in right- and left-hand circular polarizations (RCP and LCP), respectively, with an ultrahigh time cadence of 0.05 s. We performed standard flux, bandpass, and delay calibration against 3C48 and complex gain calibration against J2130+0502\footnote{3C48 is a celestial source with a well-characterized flux density spectrum, and J2130+0502 is a point-like celestial source with an accurately known position. For more details on the VLA calibration procedure, please refer to \url{https://science.nrao.edu/facilities/vla/docs/manuals/obsguide/calibration} and references therein.}. The gain changes associated with the 20-dB solar attenuators were corrected using methods described in \citet{2013PhDT.......498C}.

VLA cross-power dynamic spectrum\footnote{The cross-power dynamic spectra are obtained by doing a median of the visibilities obtained at several short baselines, which is a proxy of the total-power dynamic spectrum but contains modulations from large-scale structures (such as the horizontal stripes).} shows a strong radio QPP feature during the flare impulsive phase (Figure \ref{fig: lc}(c)). In Section \ref{sec: radio}, we will show that, with dynamic imaging spectroscopy, the radio emission during the impulsive phase actually includes not only the strong QPP, but also two other sources located at two distinctively different locations.

\subsection{Radio Imaging Spectroscopy} \label{sec: radio}
\begin{figure*}
\plotone{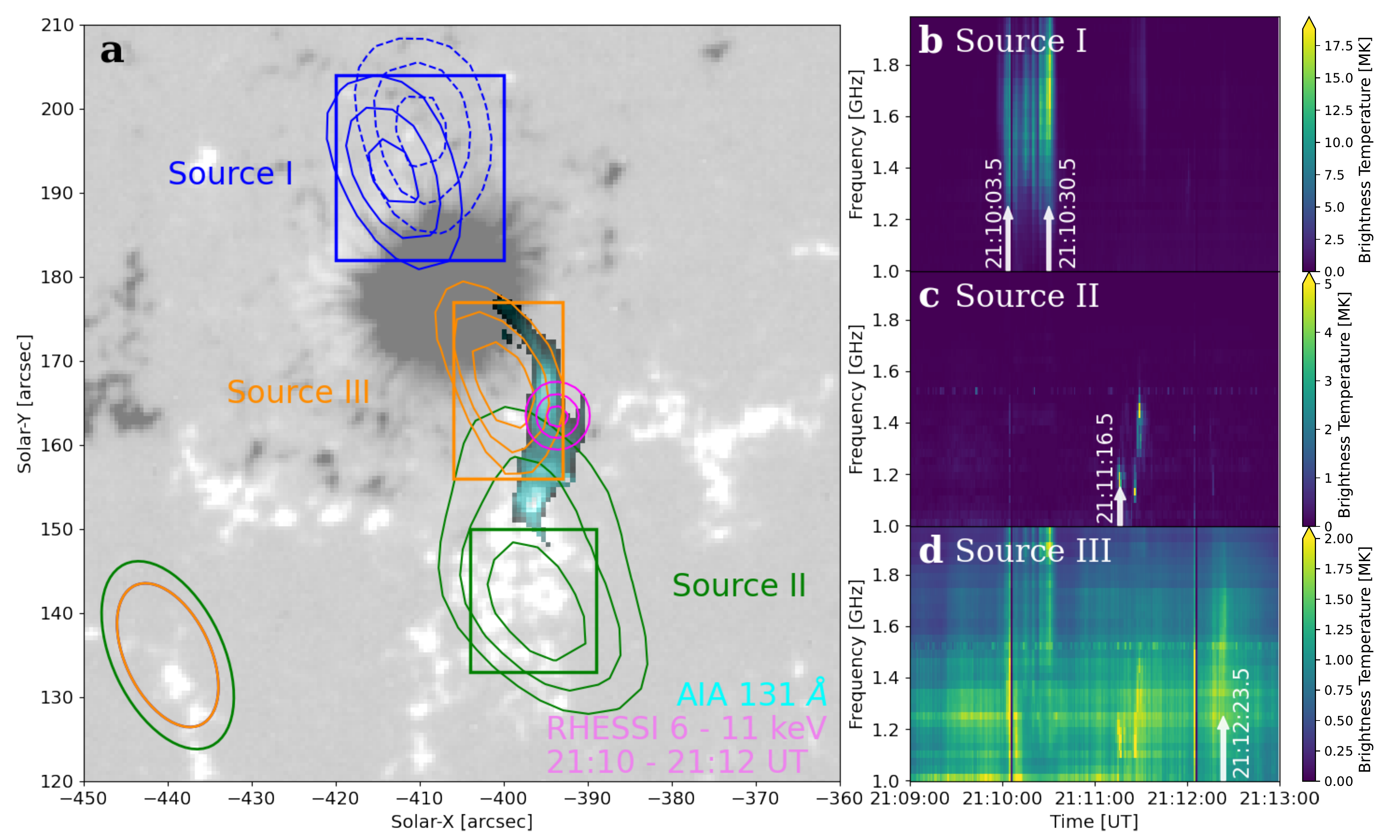}
\caption{Location of the different radio sources observed by the VLA during the flare impulsive phase and their spatially resolved dynamic spectra. (a) Radio images for the three observed radio sources (contour levels are 70\%, 80\%, and 90\% of the brightness maximum in the image). Sources I \edit1{(solid and dashed contours represent the images made close to the beginning and end of the source duration, respectively)} and II are shown in RCP and LCP, respectively. Source III is shown in Stokes I. The times used to make the images are marked as white arrows in (b)--(d). Magenta contours show the RHESSI 6--11 keV X-ray source (contour levels are 30\%, 60\%, and 90\% of keV, the maximum). The background is the SDO/HMI line-of-sight magnetogram image (gray scale) overlaid with SDO/AIA 131 \AA\ image at 21:10 UT that shows the bright flare arcade (blue shaded). The radio synthesized beams are shown in the lower left corner (orange for sources I and III at 1.5 GHz and \edit1{green} for source II at 1.18 GHz). (b)--(d) Vector dynamic spectrum of the three sources by integrating over the radio brightness over the boxes shown in (a).\label{fig: vecdy}}
\end{figure*}

With dynamic imaging spectroscopy, three spatially distinct radio sources are revealed during the flare impulsive phase. An example is shown in Figure \ref{fig: vecdy}(a). Two of them are located close to the conjugate footpoints of the erupting flux rope: one near the major sunspot (Source I, blue\edit1{ solid and dashed} contours, made at 21:10:03.5 UT \edit1{and 21:10:30.5 UT} in 1.2--1.8 GHz, respectively; RCP) and the other located near the southern ribbon (Source II, green contours, made at 21:11:16.5UT in 1.1--1.25 GHz; LCP). The third source is located at the top of the bright flare arcade (Source III, orange contours, made at 21:12:23.5 UT in 1.2--1.7 GHz, Stokes I), coinciding with the 6--11 keV X-ray looptop source observed by RHESSI (magenta contours). In order to reveal the intrinsic spectro-temporal properties of each source, we adopt the method of generating spatially resolved ``vector'' dynamic spectra: this is done by firstly imaging every time and spectral pixel, forming a four-dimensional image cube (two in space, one in frequency, one in time), and then obtaining the peak intensity within a selected region of interest that encompasses the source in the plane of the sky as a function of time and frequency \citep[see, e.g.,][]{2015Sci...350.1238C,2018ApJ...866...62C,2017SoPh..292..168M}. We obtain such vector dynamic spectra for all three radio sources, shown in Figures \ref{fig: vecdy}(b)--(d), respectively. It is evident that their spectro-temporal properties are vastly different from each other. Their main characteristics are listed in Table \ref{table: 1} and are discussed in the following.

\begin{figure*}[!ht]
\plotone{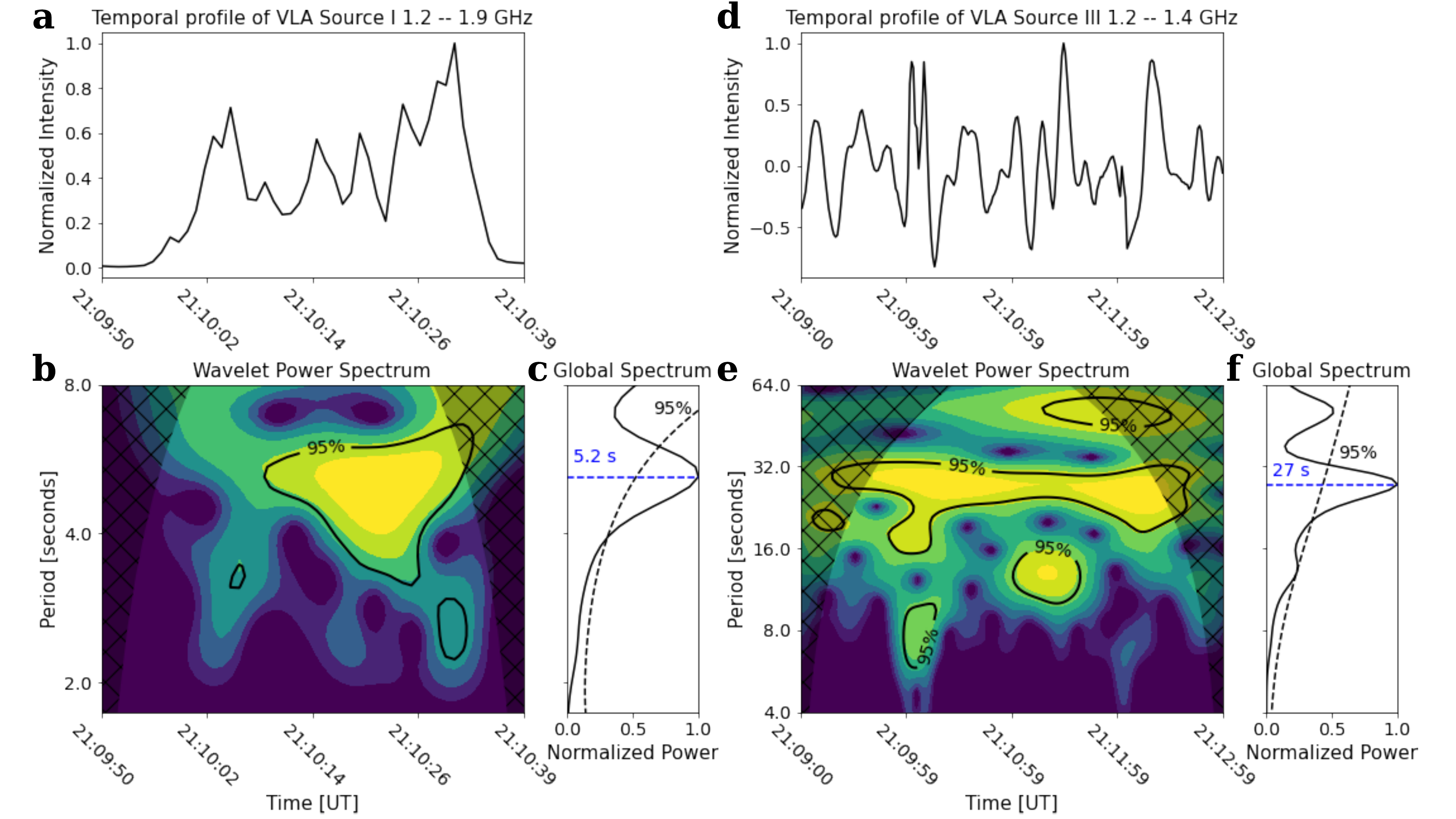}
\caption{Wavelet analysis for the two VLA radio sources that display a QPP signature (sources I and III). The left and right panels are for sources I and III, respectively. (a) shows the temporal profile \edit1{of source I averaged over 1.2--1.9 GHz and normalized by the maximum}, (b) represents the wavelet spectrum using the time profile in (a), and (c) is the global wavelet power spectrum. The solid black curves in (b) and dashed black curves in (c) display 95\% of the significance level. (d)--(f) Same as the left panels but for Source III. \label{fig: wavelet_radio}}
\end{figure*}

Source I, which is located close to the main negative sunspot, manifests as broadband QPPs that spread nearly the entire 1--2 GHz $L$ band. Several bright pulsations, which have a high brightness temperature of up to 20 MK, occur between 21:10:00 and 21:10:35 UT. They are strongly right-hand circularly polarized, suggesting that the emission is polarized in the sense of $o$-mode. The typical degree of polarization (DOP), defined as $(I_R-I_L)/(I_R+I_L)$ where $I_R$ and $I_L$ are the intensity in RCP and LCP, respectively, is 40\%--50\%, and can reach 80\% at high frequencies. \edit1{Wavelet analysis (following \citealt{1998BAMS...79...61T}) is performed on the \edit1{broadband} QPPs. The wavelet analysis result based on the light curve averaged over 1.2--1.9 GHz (Figures \ref{fig: wavelet_radio}(a)--(c)) shows that Source I QPPs have a period of around 5 s, with individual pulse lasting for $\sim$1 s.} 

Source II, located at the conjugate footpoint of the flux rope, consists of a few bright (peak brightness $\sim$8 MK), but short-lived narrowband bursts at below 1.5 GHz. Source II bursts are completely polarized in LCP. \edit1{Although Source II appears to contain multiple pulses, unlike Source I, the duration is too short to perform a periodicity analysis.} The sense of polarization and the underlying magnetic field polarity are both opposite to that of Source I, indicating that it is also polarized in the sense of $o$-mode. 

Source III, similar to Source I, is broadband in nature. However, in stark contrast to Sources I and II, it is weakly polarized (DOP $<$ 10\%) and considerably less bright (peak brightness temperature is around 2 MK). It is also the longest-lasting radio source with a duration of over 3 minutes from 21:09:30 to 21:12:30 UT. According to the wavelet analysis, its periods range from 25 to 45 s at different frequency bands. An example of the wavelet analysis results is shown in Figures \ref{fig: wavelet_radio}\edit1{(d)--(f)} for the frequency range of 1.2--1.4 GHz\edit1{, where most of the Source III pulses are observed}. The light curve used for the wavelet analysis has been averaged in 1 s and detrended to remove the slow-varying component (see \citealt{2011A&A...533A..61G,2016ApJ...825..110A} for the discussion on the detrending method, here we adopt a window of 10 s to do the smooth estimate and subtract it from the original data for detrending). Interestingly, Source III shares similar temporal characteristics to the (total-power) X-ray emission during the same period, which will be discussed in Section \ref{sec: x-ray}.

\begin{table*}[ht!]
\caption{Properties of Different Radio Sources}
\centering
\begin{tabular}{lcccccc}
  \hline
   & Location & Peak $T_b$ & Polarization (Sense \& DOP) & Period & Bandwidth & Duration \\ 
  \hline
  Source I & Footpoint & 20 MK & RCP, 40\%--80\% & 5 s & $>$1 GHz & $\sim$35 s\\ 
  \hline
  Source II & Footpoint & 8 MK & LCP, $>$90\% & N/A & $<$0.4 GHz & $\sim$25 s\\ 
  \hline
  Source III & Looptop & 2 MK & Weakly RCP, $<$10\% & 25--45 s & $>$1 GHz & $>$180 s\\ 
  \hline
\end{tabular}

\label{table: 1}
\end{table*}

\subsection{X-Ray Observations} \label{sec: x-ray}

\begin{figure*}[!ht]
\plotone{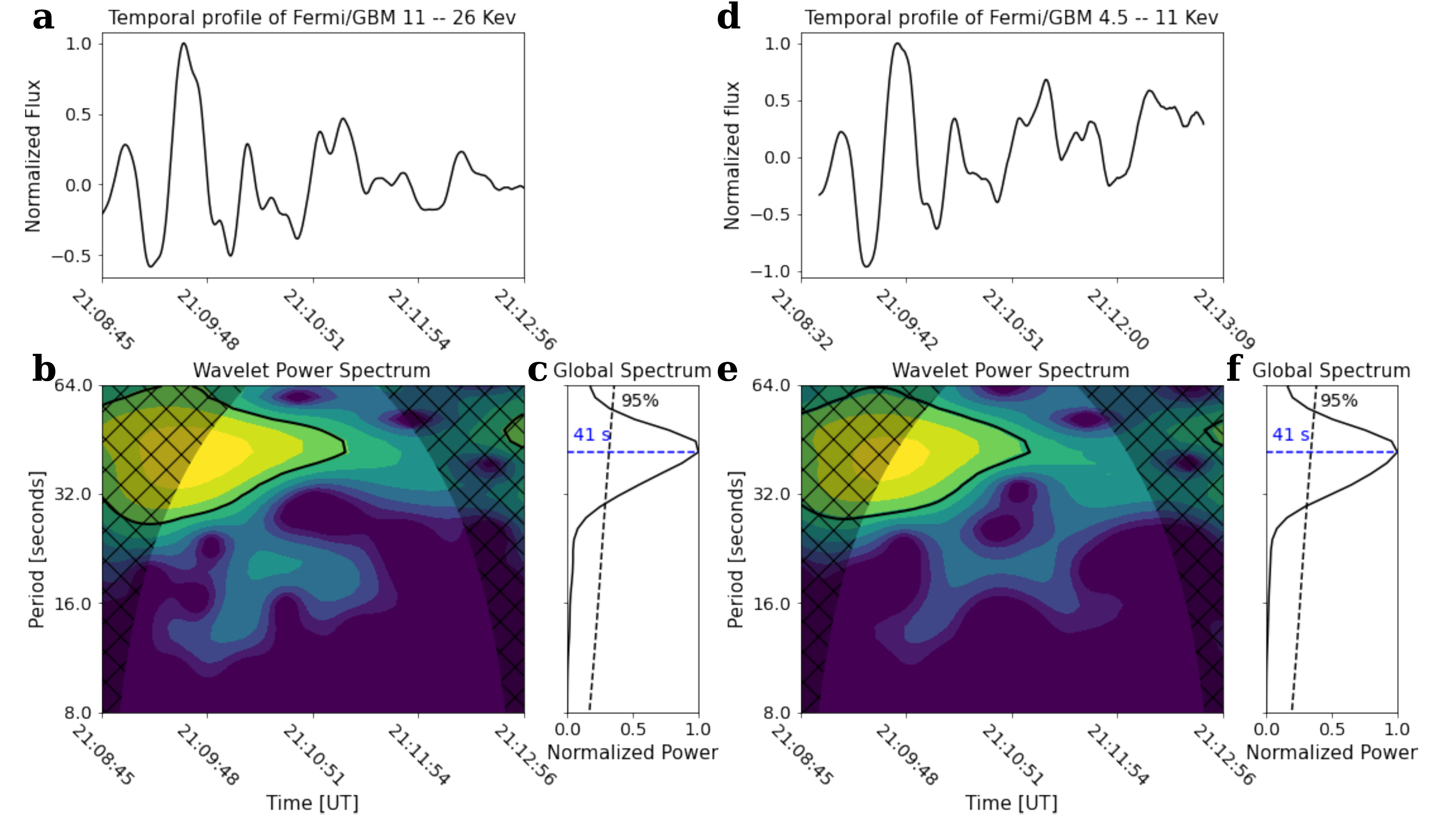}
\caption{Wavelet analysis of Fermi/GBM 11--26 keV (left panels) and 4.5--11 keV X-ray (right panels) light curves. The subplot layout is the same as in Figure \ref{fig: wavelet_radio}. All the light curves are detrended to remove the slow-varying component.\label{fig: wavelet_xray}}
\end{figure*}

\begin{figure*}[!ht]
\plotone{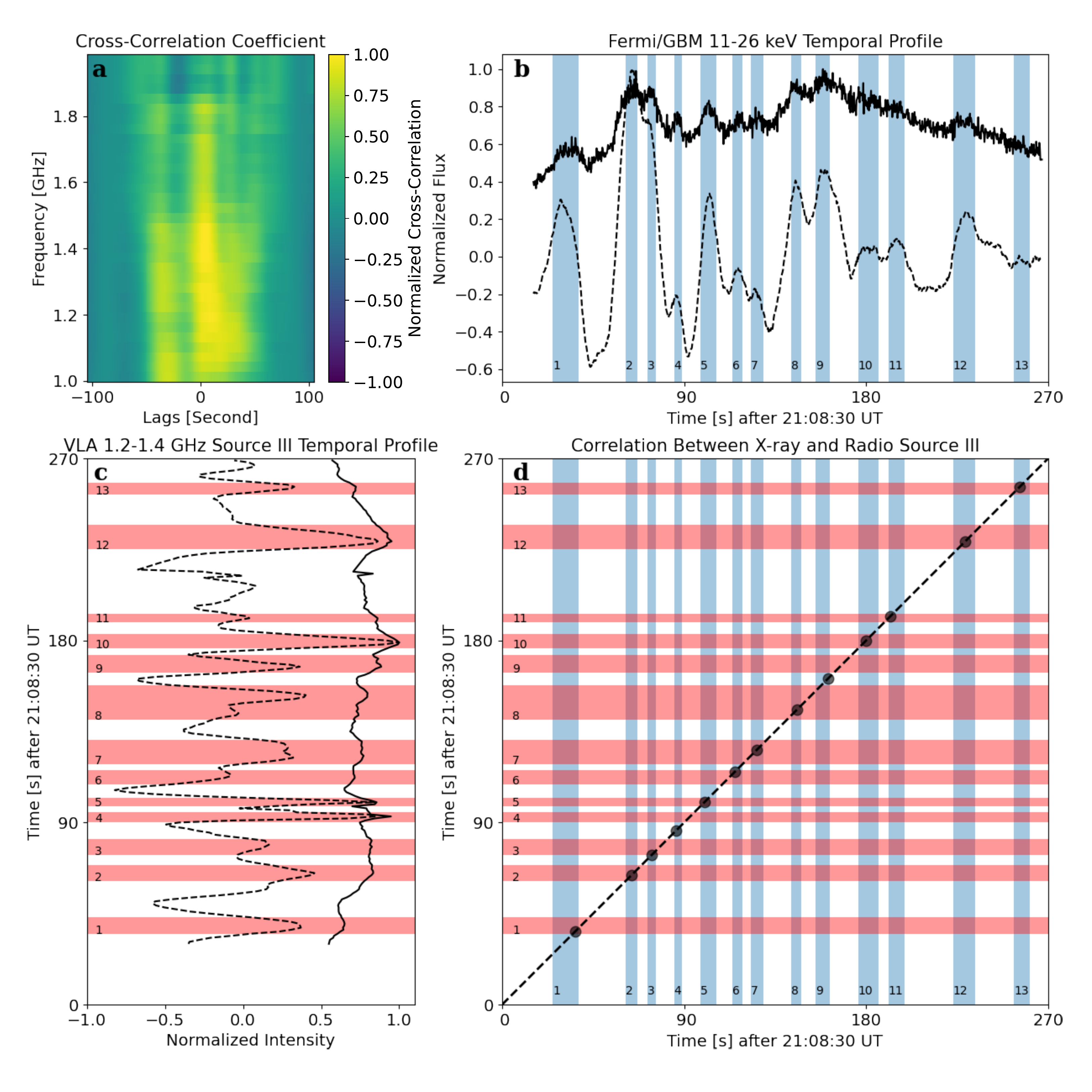}
\caption{Temporal correlation between the light curves of the VLA radio looptop source (source III) and the Fermi/GBM X-ray emission. (a) Cross-correlation coefficient between the VLA source III and Fermi/GBM 11--26 keV light curves. The cross-correlation is calculated using the detrended VLA source III vector dynamic spectrum and the Fermi/GBM 11 -- 26 keV flux temporal profile. The results are normalized to unity. (b) and (c) Identification of individual radio and X-ray pulses on Fermi/GBM 11--26 keV X-ray and VLA source III 1.2--1.4 GHz radio light curves, respectively (solid curves represent the original normalized light curve and dashed curves represent the detrended light curves). The FWHM duration of each pulse is marked in thick bars. (d) Correlation plot between the times of the X-ray pulsations and those of the radio pulsations. The horizontal (blue) and the vertical (red) thick bars mark the times of the X-ray pulses and those of the radio pulses, respectively. \label{fig: corr}}
\end{figure*}

During the impulsive phase of the flare, quasiperiodic X-ray bursts are observed by RHESSI and Fermi/GBM. RHESSI 6--11 keV image shows that the X-ray source is located at the looptop region, coinciding with the VLA radio source III (shown in Figure \ref{fig: vecdy}(a) as magenta and orange contours, respectively). Wavelet analysis of the detrended X-ray light curve shows a period of $\sim$41 s (Figure \ref{fig: wavelet_xray}). In addition to their spatial coincidence, the X-ray and radio looptop sources also share a temporal correlation. At each radio frequency, a cross-correlation analysis is performed between the VLA radio source III light curve and the Fermi/GBM 11--26 keV light curve. The light curves of both sources are averaged in 1 s and detrended with a smoothing window of 10 s (dashed curves in Figures \ref{fig: corr}(b) and (c)). The cross-correlation coefficient as a function of time lags and frequency is shown in Figure \ref{fig: corr}(a). The peak of the cross-correlation coefficient lay within $\pm$5 s, or $<$1/6 of the period of the QPPs, indicating that they are nearly simultaneous. To further demonstrate their temporal correlation, following the method used by \citet{2001ApJ...562L.103A}, we also identify every X-ray and radio pulse using a Gaussian decomposition method. Each burst is identified along with a half-width of the local peak shown as the shadowed area in Figures \ref{fig: corr}(b) and (c). A consistent check is also performed using the original, non-detrended light curves (solid curves in Figures \ref{fig: corr}(b) and (c)). As shown in the correlation plot in Figure \ref{fig: corr}(d), the identified pulses from radio and X-ray show striking similarities: almost every X-ray pulse has its counterpart in radio. Such a temporal similarity strongly suggests that the X-ray and radio QPP are associated with the same energetic electron population \citep{2001ApJ...562L.103A,2003ApJ...588.1163G}. We note that the wavelet-derived periods of X-ray QPPs are slightly longer than that of the radio source III. The difference may be attributed to the lack of contribution of the weak bursts to the wavelet analysis results (for example, the X-ray bursts numbered 4, 7, and 10 in Figure \ref{fig: corr}(b)).

\begin{figure*}
\plotone{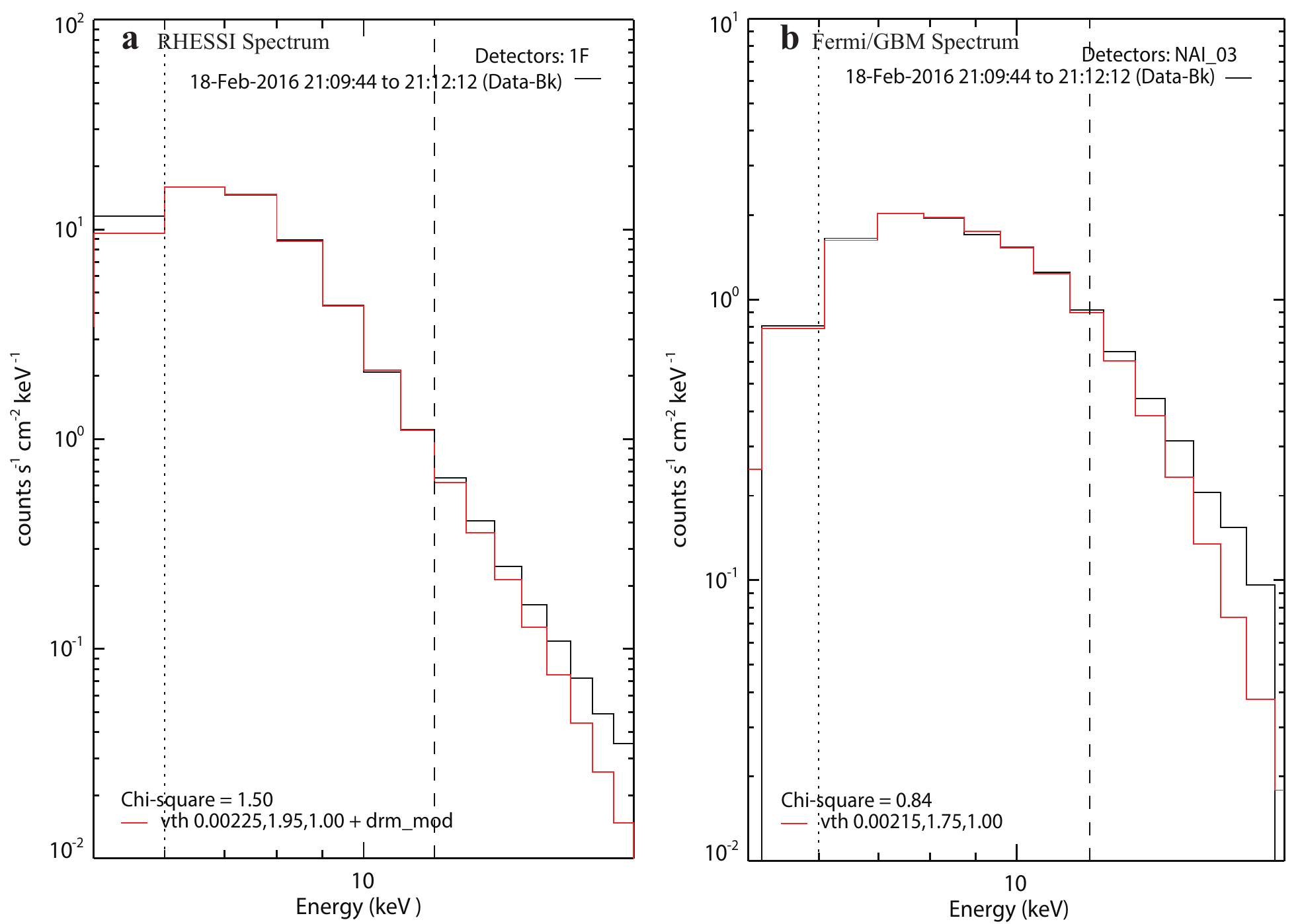}
\caption{RHESSI and Fermi/GBM X-ray spectra and fit results. (a) RHESSI X-ray spectrum obtained by its detector 1 for 21:09:44--21:12:12 UT (black curve). \edit1{We use an isothermal model (``vth'') to fit the observed spectrum in 6--12 keV (demarcated by the dotted and dashed lines, respectively). The higher-energy range is excluded due to the impact of the pulse pileup. Also included in the fit is the  ``drm-mod'' module for fine-tuning the RHESSI response (not shown). The fit returns a volume emission measure of $0.00225\times10^{49}$ cm$^{-3}$, and a plasma temperature of 1.95 keV or 22.6 MK. The relative coronal abundance parameter is fixed to 1.0. The reduced chi-squared value of the fit is 1.50. } (b) Similar to (a), but showing Fermi/GBM X-ray spectrum (black curve) and the corresponding isothermal fit results (red curve) based on data from its detector 3 for the same time interval as RHESSI. \edit1{The best-fit parameters and the chi-square value are indicated in the legend.}
}.
\label{fig: xray fit}
\end{figure*}

We also performed spectral analysis of the X-ray data using the \texttt{OSPEX} package. Figures \ref{fig: xray fit}(a) and (b) show the RHESSI and Fermi/GBM flare-integrated X-ray spectrum obtained from 21:09:44 to 21:12:12 UT. We select 21:05:00 to 21:06:00 UT prior to the flare as the background time. An isothermal fit is performed for both RHESSI and Fermi/GBM between 6--12 keV. A clear excess of the X-ray count flux above the thermal component is present at above 12 keV in both the RHESSI and Fermi/GBM spectra, suggestive of the presence of a nonthermal component of the looptop X-ray source. However, we caution that the RHESSI spectrum above 12 keV is probably affected by the pileup effect according to the results from \texttt{hsi\_pileup\_check.pro} in \texttt{SSWIDL}. It is unclear whether Fermi/GBM also suffers from the pileup effect. Therefore, a quantitative analysis of the spectra at above 12 keV is difficult.

\section{Radio QPPs: Emission Mechanisms} \label{sec: analysis}
In Section \ref{sec: observations}, we have obtained the vector dynamic spectrum for each radio source located at different regions in the flare. In this section, we will discuss their emission mechanisms based on their spatial, temporal, spectral, and polarization properties.

\subsection{Conjugate Footpoint Sources} \label{sec: ECME}

As presented in Section \ref{sec: observations}, sources I and II are located at the conjugate footpoints of the erupting flux rope. They are highly polarized in the opposite sense. Their relatively high brightness temperature and polarization degree are both suggestive of a coherent emission mechanism. Two coherent emission mechanisms are usually relevant to decimetric radio bursts in solar flares: plasma radiation and electron cyclotron maser (ECM) emission. The former is usually expected in the corona where $\nu_{\rm pe} \gg \nu_{\rm ce}$ ($\nu_{\rm pe}$ is the plasma frequency and $\nu_{\rm ce}$ is the electron cyclotron frequency) and the latter is usually relevant in the region with a relatively strong magnetic field which satisfies $\nu_{\rm ce} \gtrsim \nu_{\rm pe}$) \citep{1985ARA&A..23..169D,1998ARA&A..36..131B}.

\begin{figure*}
\plotone{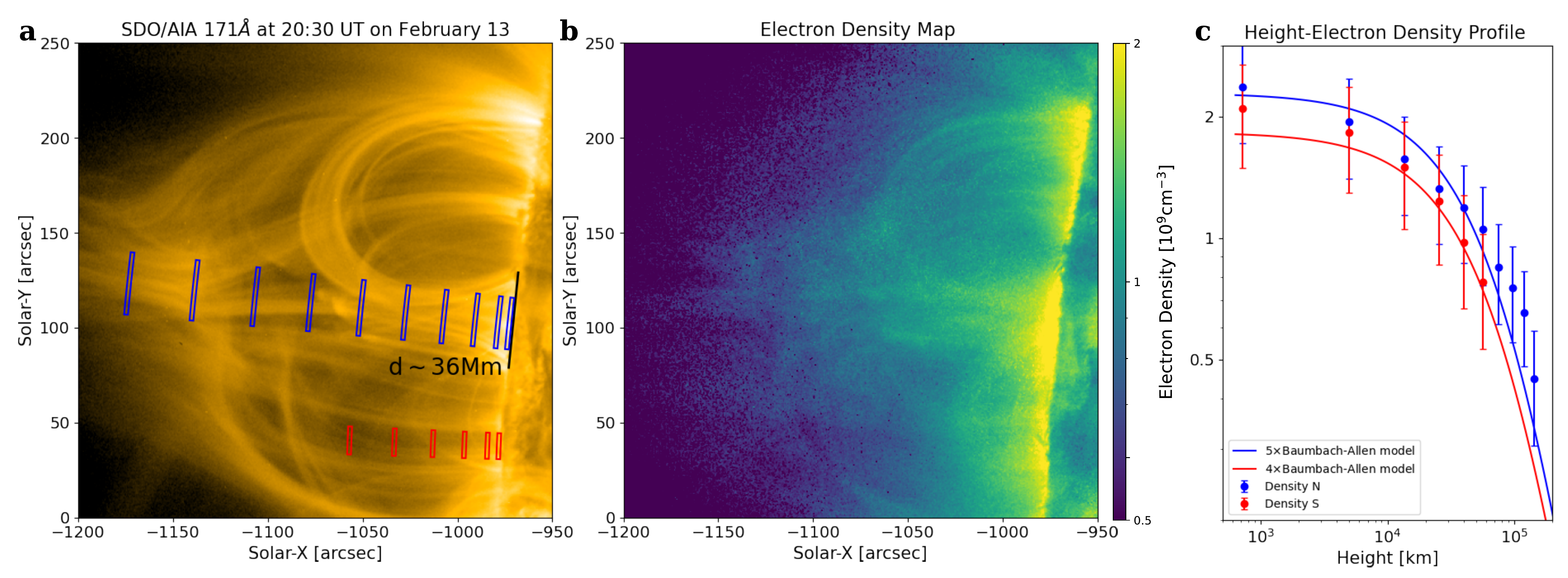}
\caption{Coronal electron density profile of the active region derived from limb observations on 2016 February 13. (a) SDO/AIA 171 \AA\ image at 20:30 UT on 2016 February 13, which features a limb-view perspective of the target active region. Column depth in generating the electron density map is taken from the coronal loop width marked with a black line segment ($\sim$36 Mm). (b) Electron density map derived from DEM analysis. (c) Derived electron number density with height from regions marked in (a) for a series of radial heights above the main negative sunspot (blue) and the southern positive polarity (red). The blue and red curves are fit to the DEM-derived coronal density using five and fourfold of the Baumbach–Allen density model.
\label{fig: cor_den}}
\end{figure*}

\begin{figure*}
\plotone{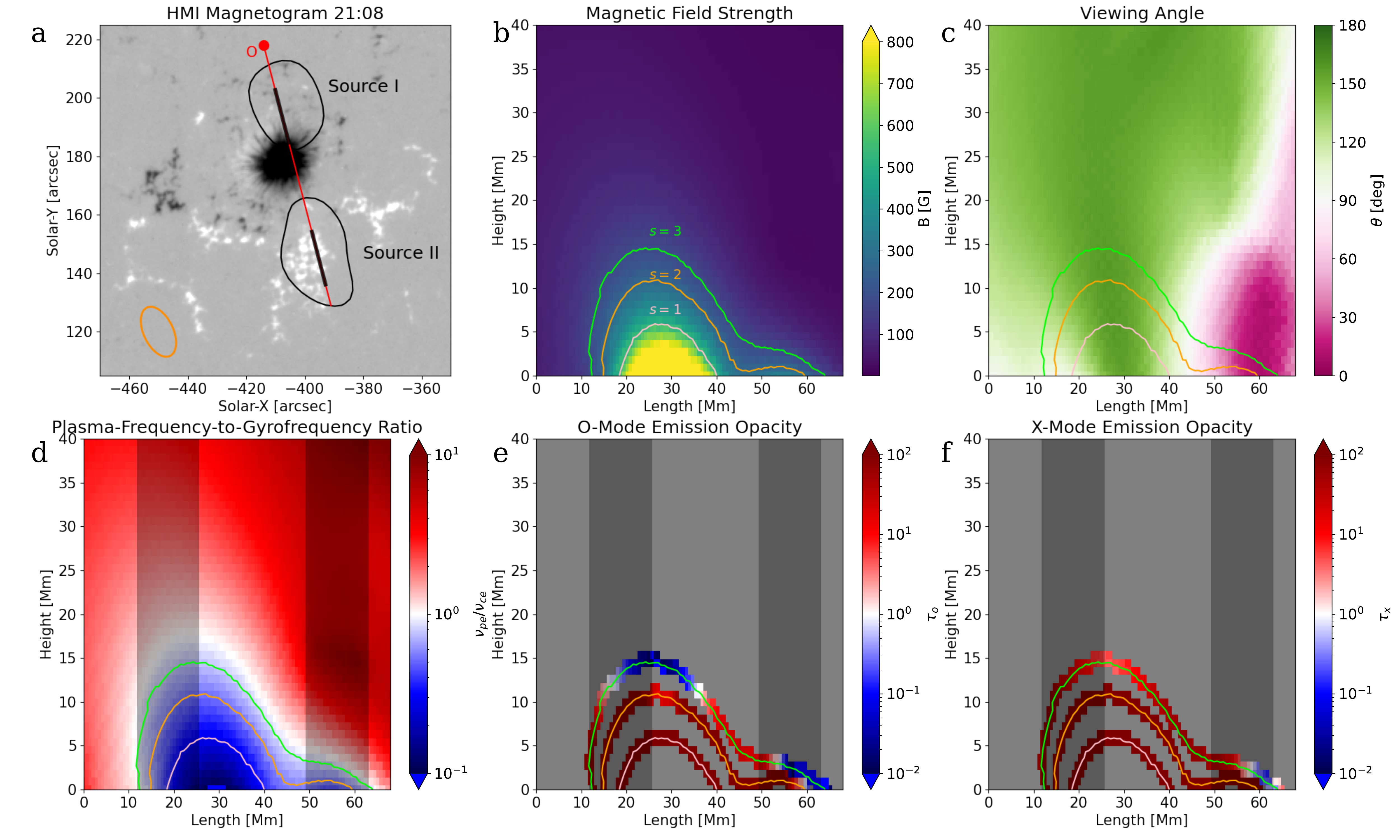}
\caption{Coronal parameters at the locations of radio sources I and II. (a) Context image of radio sources I (50\% contour) and II (70\% contour) overlaid on SDO/HMI line-of-sight magnetogram. (b) The magnetic field strength in a selected two-dimensional plane. The bottom boundary of the plane is shown as the red line in (a) and its vertical axis is along the line of sight. Pink, orange, and green contours represent the first, second, and third harmonics of the electron gyrofrequency $\nu_{\rm ce}$ for a representative emission frequency $\nu$ at 1.5 GHz, respectively (i.e., $\nu=s_{\rm{e}}\nu_{\rm ce}$, where $s_{\rm{e}}=1$, 2, 3). (c) Similar to (b), but showing the viewing angle between the local magnetic field vector and the line-of-sight direction $\theta$. (d) Distribution of the plasma-frequency-to-gyrofrequency ratio $\Omega=\nu_{\rm pe}/\nu_{\rm ce}$ in the selected plane. Blue and red regions are for $\Omega < 1$ and $\Omega > 1$ regions, respectively. The two shaded regions represent the line-of-sight locations of sources I and II (the width is the same as the synthesized beam size shown as an orange ellipse in (a)). (e) and (f) Calculated opacity $\tau_{o, x}$ at the first, second, and third harmonic layers for the $o$- and $x$-mode, respectively.
\label{fig: ecme_ck}}
\end{figure*}

To evaluate the relative importance of plasma radiation and ECM emission at the source site, we adopt the coronal magnetic field derived from the NLFFF extrapolation results as the magnetic field model. Constraining the plasma density distribution as a function of height, however, is not straightforward for this event owing to its on-disk viewing perspective. Fortuitously, the magnetic topology of the AR shows no significant change since its first appearance on the east limb on 2016 February 13, five days before the event under study. Therefore, we use the plasma density profiles derived using a differential emission measure (DEM) method \citep{2012A&A...539A.146H} based on SDO/AIA multi-band EUV imaging data of the limb-view AR obtained on 2016 February 13 as a proxy for our analysis. The results are shown in Figure \ref{fig: cor_den}(b). To estimate the density from the DEM results, a uniform column depth of 36 Mm is assumed \edit1{with a nominal uncertainty of 50\%}. The value is taken as the width of the coronal loop bundles seen in the AIA images. Furthermore, we derive the density profile by taking the average of the density values in a series of transverse slices (shown in Figure \ref{fig: cor_den}(a)) placed at different radial heights above sources I and II. The results are shown in Figure \ref{fig: cor_den}(c) as blue and red symbols, respectively. The uncertainties of the density values at a given height are calculated according to those from the DEM results and column depth through standard error propagation methods. 
We then use the Baumbach–Allen density model \citep{1937AN....263..121B,2000asqu.book.....C}: 
\begin{equation}
    n_e(R)=C\left(2.99R^{-16}+1.55R^{-6}+0.036R^{-1.5}\right)\times 10^8\ \rm{cm}^{-3},\label{bau-allen}
\end{equation} 
where $R$ is the radial distance from the solar center expressed in solar radius $R_{\odot}$, and $C$ is the scaling factor, to fit the DEM-constrained density profiles. We find that the four- or five-fold Baumbach–Allen density model (i.e., $C$=4 or 5) yields the best fit to the data (Figure \ref{fig: cor_den}(c)). In the following analysis, for simplicity, we will adopt the five-fold Baumbach–Allen model and apply it as the general density model uniformly across the AR (the fourfold model returns very similar results).

\begin{figure*}
\includegraphics[width=1.0\textwidth]{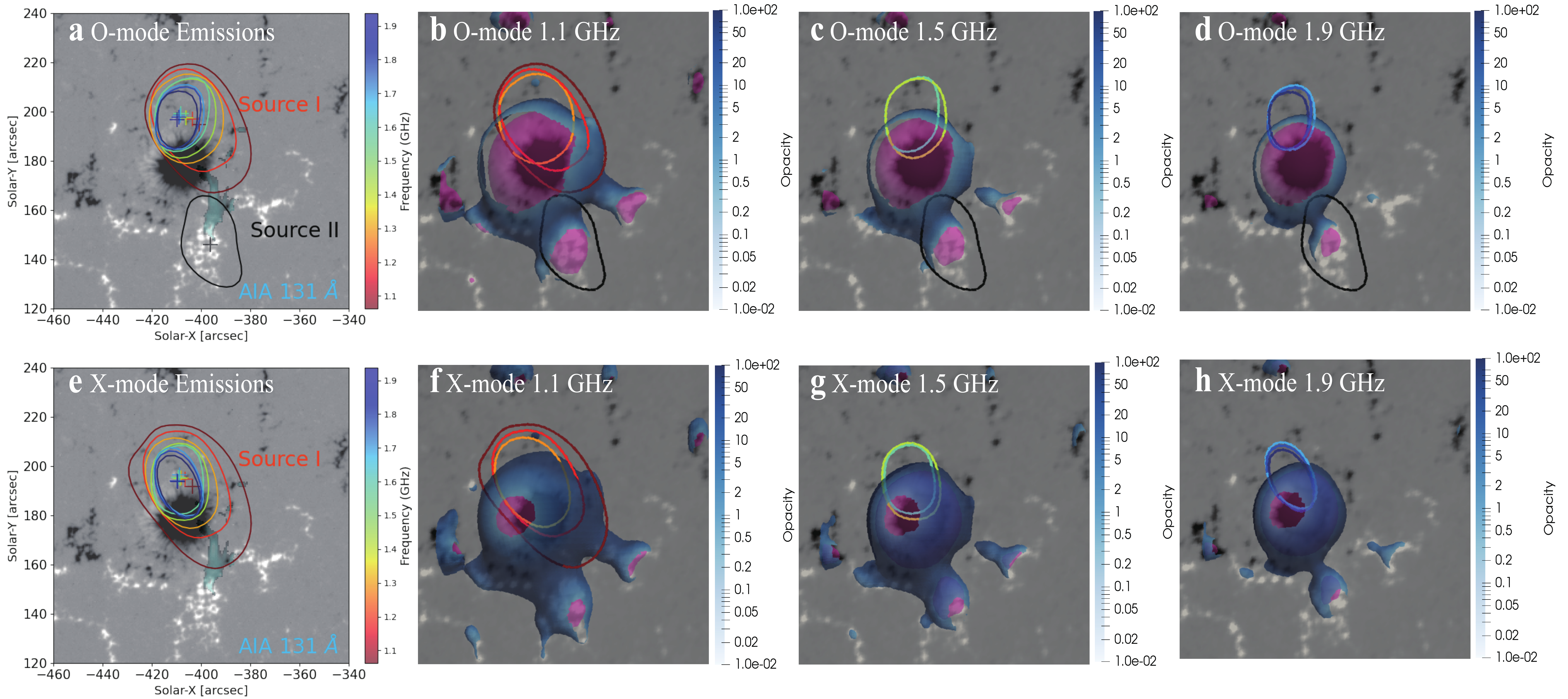}
\caption{(a) Radio $o$-mode sources I and II observed by VLA at multiple frequencies. Color contours are 50\% contours for source I, and their centroids are marked with a plus sign. The black contour is the 70\% contour of source II at 1.2 GHz. (b)--(d) Calculated gyroresonance opacity $\tau_{o, x}$ at the third harmonic layers for 1.1 GHz, 1.5 GHz, and 1.9 GHz (i.e., $\nu = s_{\rm{a}}\nu_{ce}$, where $s_{\rm{a}}=3$), respectively. All panels show the view along the line of sight. The opacity at the gyroresonance layer is colored in blue and the region with an opacity smaller than unity is set to be transparent. The second harmonic gyroresonance layer is shown as a magenta surface underneath the third harmonic layer. The observed VLA radio sources are also overlaid as color contours. (e)--(h) Same as the upper panels but for the $x$-mode.\label{fig: ecme}}
\end{figure*}

With both the magnetic and density model, we can now calculate the ratio of the plasma frequency to the electron cyclotron frequency $\Omega=\nu_{\rm{pe}}/\nu_{\rm{ce}}$ at any point of interest in the corona. Figure \ref{fig: ecme_ck}(d) shows the distribution of $\Omega$ in a vertical plane that passes the centroids of radio sources I and II near the footpoints of the erupting flux rope. It can be clearly seen that $\Omega < 1$, the condition favoring the ECM mechanism, is satisfied above the sunspot region at relatively low coronal heights ($\lesssim$10 Mm) owing to its strong magnetic field. In particular, this $\Omega < 1$ region encompasses the entire volume where low harmonics of electron gyrofrequency is located for our observing frequency (color contours in Figure \ref{fig: ecme_ck}(b), which show the first, second, and third harmonic layers with $[\nu/\rm{Hz}]\approx2.8\times10^{6}s_{\rm{e}}[B/\rm{G}]$, where $s_{\rm{e}}$ is the harmonic number of the ECM emission from the source region).  Hence, if radio sources I and II are due to ECM emission at low harmonics of the electron gyrofrequency, the local conditions would favor the needed wave growth. 
In comparison, plasma radiation at the observed frequency of 1--2 GHz requires an electron number density of 1.2--$5\times10^{10}\,\rm{cm}^{-3}$, which corresponds to a very low height (beyond the maximum of DEM-derived electron density, $\sim\!3\times10^{9}\,\rm{cm}^{-3}$), entering the transition region/upper chromosphere. Unless a strong over-dense region exists in the otherwise tenuous corona (see, however, an example reported by \citealt{2013ApJ...763L..21C}), plasma radiation is deemed unlikely to account for the observed sources I and II. 

Furthermore, the observed locations of sources I and II are both near the regions where the magnetic field lines converge, which is favorable for ECM emission generated due to the loss-cone instabilities \citep{1990A&A...229..206A}. In addition, the ECM emission, generated at low harmonics of the electron cyclotron frequency, needs to overcome the absorption by the overlying gyroresonance layers with a magnetic field that corresponds to a higher harmonics of the electron gyrofrequency at $\nu=s_{\rm{a}}\nu_{\rm{ce}}=2.
8\times10^{6}\ \rm{Hz}\ s_{\rm{a}}[B/\rm{G}]$, where $s_{\rm{a}}>s_{\rm{e}}$ is the harmonic number of the absorbing gyroresonance layer. To investigate the absorption effect, at a given observing frequency $\nu$, we first locate the corresponding gyroresonance isoGauss layers at different harmonics of $s_{\rm{a}}$ with a magnetic field strength that satisfies $[\rm{B}/\rm{G}]\approx[\nu/\rm{Hz}]/2.8\times10^{6}s_{\rm{a}}$. We then calculate the optical depth $\tau_{x,o}(s, \nu, \theta)$ at these gyroresonance layers for both the $x-$ and $o-$ mode using the following equation (after \citealt{gary2004solar}):
\begin{equation}
    \tau_{x,o}(s,\nu,\theta)=0.0133\frac{n_eL_B(\theta)}{\nu}\frac{s^2}{s!}(\frac{s^2\sin^2\theta}{2\mu})^{s-1}F_{x,o}(\theta)\label{equtau}
\end{equation} 
where $\theta$ is the angle between the line of sight (LOS) and the magnetic field direction (taking the acute angle, ranges [0$^{\circ}$, 90$^{\circ}$]), $L_B(\theta)$ is the scale length of the magnetic field along the line of sight ($L_B(\theta)=B/\frac{dB}{dl}$). $F_{x,o}(\theta)$ is a function of angle which can be approximated as (for low harmonics and $\theta$ away from 90$^{\circ}$): 
\begin{equation}
    F_{x,o}(\theta)\approx(1-\sigma cos\theta)^2,
\end{equation} 
where $\sigma=\pm1$ for the $o$- and $x$- mode, respectively. By comparing the source location and opacity of the gyroresonance layers at different harmonics, we conclude that the second harmonic ($s_{\rm{e}}=2$) ECM emission is the most likely for the emission to escape from the overlying gyroresonance layers with a higher harmonics (i.e., $s_{\rm{a}}>2$). More specifically, for the fundamental ECM emission ($s_{\rm{e}}=1$), the opacity at the second harmonic gyroresonance layer ($s_{\rm{a}}=2$) is very large at all locations above the source site (see Figures \ref{fig: ecme_ck}(e) and (f)), while for ECM at higher harmonics ($s_{\rm{e}}>3$), the maser growth is much less efficient \citep{2002ApJ...575.1094W,2013JGRA..118.7036L}. 

We show the observed source locations (color contours) overlaid on the second-harmonic isoGauss layers (magenta domes) in Figures \ref{fig: ecme}(b)-(d), (f)-(g), with the assumption that the ECM emission is produced at the second harmonic (i.e., $s_{\rm{e}}=2$, or $[\nu/\rm{Hz}] = 2\nu_{\rm{ce}}=5.6\times10^6 \rm{[B/G]}$). We show these layers at different emission frequencies at 1.1, 1.5, and 1.9 GHz, which correspond to a magnetic field strength of $B=196$ G, 268 G, 339 G, respectively). We then ``paint'' the isoGauss domes with the calculated third-harmonic gyroresonance opacity $\tau_{o,x}(s_{\rm{a}}=3)$ along the line of sight using a white-to-blue color scale (blue means a larger opacity). The region where $\tau_{o,x}(s_{\rm{a}}=3)<1$ is rendered transparent, representing an ``opacity hole.'' It is evident from Figures \ref{fig: ecme}(b)--(d) that, for the $o$-mode, a large opacity hole is present in the close vicinity of the observed radio sources I and II, allowing the $o$-mode ECM emission to escape easily. In comparison, such an opacity hole is much smaller for the $x$-mode as shown in Figures \ref{fig: ecme}(f)--(h), causing most of the $x$-mode ECM emission to be absorbed by the overlying $s_{\rm{a}}=3$ gyroresonance layer. We argue that this is the main reason why we have observed dominating $o$-mode emission for the conjugate sources I and II from opposite magnetic polarities. \edit1{Here we note that since the radio ECM source is probably point-like and unresolved, it can be anywhere in the opacity hole (including, e.g., its edge). Therefore, the observed radio source morphology shown in Figure \ref{fig: ecme} does not necessarily match that of the opacity holes. Also, there are likely uncertainties associated with the coronal magnetic field extrapolation and the assumed density model, both of which would affect the exact shape and location of the opacity holes. Hence we conclude that the results are consistent with our proposed scenario within uncertainties.}

It is intriguing to note that while the northern radio source I is broadband, the southern source II is narrower in frequency and shorter in duration. Although fully understanding such a difference requires a more detailed radiation modeling, we suggest that they may be affected by the much smaller opacity hole above source II, rendering the emission more difficult to escape. In addition, owing to the much smaller photospheric field strength at the location of source II, ECM emission at the same frequency is likely originated from a much smaller coronal height than Source I. The local plasma frequency may be high enough to suppress, if not quench, the ECM emission. These factors may contribute to the more intermittent and narrowband nature of source II.

\begin{figure*}
\plotone{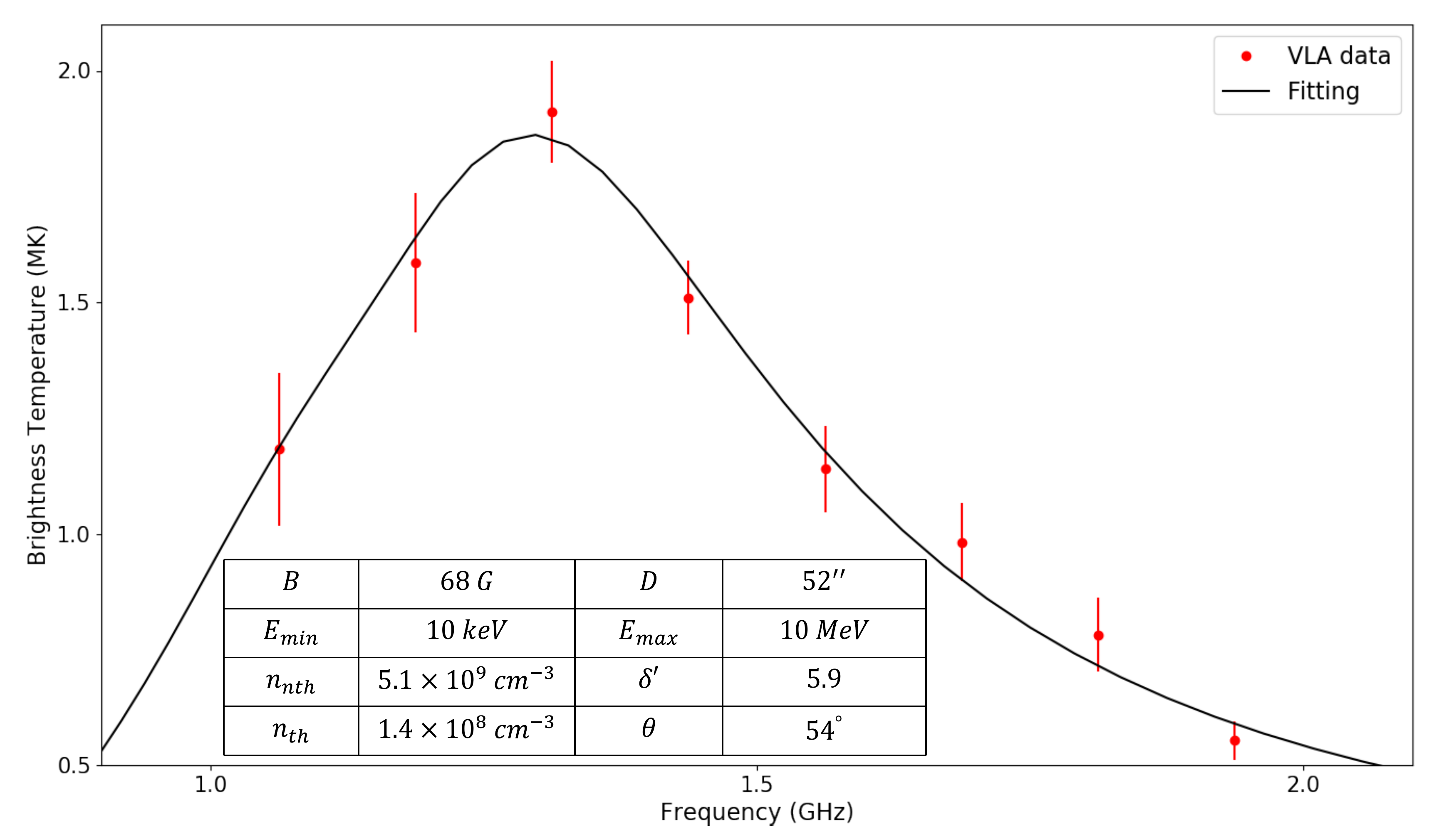}
\caption{Brightness temperature spectrum of the looptop radio source III and gyrosynchrotron fitting results. VLA data points are selected from all eight spectral windows spanning the 1--2 GHz $L$ band and the black curve is the best-fit spectrum. Also listed are the parameters of the fit spectrum, which include magnetic field strength $B$, column depth $D$, low- and high-energy cutoff of the power-law nonthermal electron distribution $E_{\rm low}$ and $E_{\rm high}$, total nonthermal electron number density $n_\mathrm{nth}$ above $E_{\rm low}$, power-law index $\delta'$ of the electron density distribution, total thermal electron number density $n_\mathrm{th}$, and viewing angle $\theta$. Note that since the source is under-resolved by VLA in this frequency range, the observed spectrum likely deviates from the assumption of a homogeneous source. Hence the fit parameters should be treated as representative values.}.
\label{fig: gs fit}
\end{figure*}

\subsection{Flare Arcade Source}\label{sec: gs}

Unlike sources I and II, radio source III is weakly polarized in RCP ($<10$\%) with a relatively low brightness temperature ($<$2 MK). We extract the average brightness temperature spectrum of the source from the VLA images made between \edit1{21:12:15--21:12:25} UT when sources I and II are temporally absent. The obtained brightness temperature spectrum of the full 1--2 GHz band (eight spectral windows) is shown in Figure \ref{fig: gs fit}. The uncertainty of each brightness temperature measurement is estimated as the root mean square variation of an empty region in the image that contains no apparent radio source. The derived radio spectrum shows a typical nonthermal gyrosynchrotron emission pattern with an optically thick, rising slope at low frequencies and an optically thin, descending slope at high frequencies. To further confirm the gyrosynchrotron nature of source III, we perform a spectral fitting using the fast gyrosynchrotron code \citep{2010ApJ...721.1127F} to calculate the model gyrosynchrotron spectrum. The best-fit spectrum is shown as a solid curve in Figure \ref{fig: gs fit} with the corresponding fit parameters listed therein. We note that our spectral analysis \edit1{is performed in a limited frequency range and} assumes a homogeneous emission source. Given the limited spatial resolution at the low frequencies, Source III is likely under-resolved (see Figure \ref{fig: vecdy}(a) for beam size versus source size), such an assumption is likely invalid. Therefore, the fit parameters used here may not be unique and may not precisely represent the exact properties of the source region. However, we argue that our fitting practice is sufficient to confirm that source III is most likely associated with gyrosynchrotron emission generated by flare-accelerated nonthermal electrons at the looptop. We note that the gyrosynchrotron nature of source III is further corroborated by its presence at the location of the flare arcade in the close vicinity of a looptop X-ray source observed by RHESSI, shown in Figure \ref{fig: vecdy}(a).

\section{Discussion and Conclusions} \label{sec: conclusion}

In the previous sections, we have reported radio and X-ray sources observed during the impulsive phase of a C1.8 flare on 2016 February 18. VLA's radio imaging spectroscopy capability allows us to identify the radio source as three spatially distinct sources. Radio source I is located close to the main negative sunspot near the northern footpoint of the erupting magnetic flux rope. Radio source II is present near the conjugate southern footpoint of the flux rope above a positive magnetic polarity. They both have a strong polarization with opposite senses, suggestive of $o$-mode radiation. Source III, unlike sources I and II, is weakly polarized and less bright. It shares a similar spatial location with an RHESSI X-ray source at the top of the flare arcade.

We then investigate the emission mechanism responsible for the different radio sources based on their location, polarization, and spectral properties. NLFFF extrapolation is performed to construct the magnetic field strength and direction in the AR. By comparing with the plasma frequency using a DEM-derived density model, we argue that sources I and II are likely due to the coherent ECM emission. In addition, the sense of polarization (in $o$-mode) is consistent with the gyroresonance opacity above the two sources. For radio source III, we performed spectral analysis and concluded that it is likely due to incoherent gyrosynchrotron emission from flare-accelerated nonthermal electrons trapped in the looptop region. The same energetic electrons are likely responsible for the presence of the X-ray looptop source, which shares a similar spatial location and has a close temporal correlation to this radio source. Two of the broadband radio sources---sources I and III---exhibit a quasiperiodic pattern. In the following, we will discuss their oscillation mechanisms.

Radio source I has a relatively short period of $\sim$5.2 s. We have shown that it is most likely associated with the coherent ECM emission driven by nonthermal electrons. Because of the complex emission processes that involve electron injection, nonlinear wave growth, and conversion, the modulation of the emission could occur at any stage. As introduced previously, MHD oscillations are one of the plausible drivers of the QPP. Sausage mode MHD oscillations in coronal conditions have been suggested to be responsible for second- to ten-second-scale QPP events \citep{2004ApJ...600..458A,2011ApJ...729L..18M,2011ApJ...740...90V,2013ApJ...777..159Y,2019NatCo..10.2276C}. The period of such oscillation is \citep{2012ApJ...761..134N}: 
\begin{equation}
    P_{sau} \lesssim 2.62\frac{a}{\sqrt{v_A^2+c_s^2}},\label{equsaup}
\end{equation} 
where $a$ is the coronal loop cross-section radius, $v_A$ is the Alfv\'{e}n speed and $c_s$ is the sound speed. In this event, we have shown that the source region has a relatively strong magnetic field strength (179--357 G). With a plasma density of $5\times10^{8}$--$5\times10^{9}$ cm$^{-3}$, the observed period of 5.2 s of source I requires a magnetic tube with a radius of 10--61 Mm to host the oscillations, which is comparable with the cross-section of the EUV loop bundles as observed by SDO/AIA (Figure \ref{fig: cor_den}). Other types of MHD oscillation such as the kink mode usually have much longer periods than those reported here \citep{2021SSRv..217...66Z}. \edit1{One possible scenario of ECM emission modulated by sausage mode MHD oscillations was suggested by \citet{1990SoPh..130..151Z}, in which the sausage mode can modulate the magnetic field configuration and the pitch angle distribution of the energetic electrons in the coronal loop. The modulation can subsequently result in a quasiperiodic variation of the loss-cone instability and, consequently, the ECM emission.} 

For short-period QPPs associated with coherent radio emissions, another candidate is self-organized wave-wave or wave-particle interactions \citep{1988ApJ...332..466A,1994SoPh..153..403F,1998R&QE...41...28K,2007SoPh..246..431C}. The periodicity arises as the system tries to balance itself between wave growth and wave-particle diffusion. The oscillation periods can be estimated as $\tau_p=2\pi \sqrt{\tau_{\rm{growth}}\times|\tau_{\rm{diff}}|}$ when the system is close to the steady state \citep{1988ApJ...332..466A}. Here $\tau_{\rm{growth}}$ and $\tau_{\rm{diff}}$ are characteristic times for wave growth and diffusion, respectively. Although we do not have sufficient information to constrain these time scales, the observed period falls into a plausible range in coronal conditions \citep{1988ApJ...332..447A,2011A&A...526A.161K}.

Quasiperiodic injections of energetic electrons into \edit1{the ECM source region with a loss-cone type distribution} is another possibility. Freshly injected particles can temporally fill up the loss cone to reduce the instability of the ECM wave growth, which can subsequently reduce, or even quench, the ECM emission \citep{1975A&A....45..135Z,1976SoPh...46..275B,1994SoPh..153..403F,2008ApJ...689.1412C}. This process will lead to a quasiperiodic modulation of the emission intensity and may account for the observed QPP behavior of radio source I. However, despite having an ultrahigh time resolution of 0.05 s, we did not detect any frequency drift in the source I pulses, which is a typical feature of the injection of fast electron beams in the dynamic spectrum. Assuming ECM emission, the expected frequency drift can be written as: $\left|\dot{\nu}\right|/\nu=v_b/L_B$, where $v_b$ is the speed of the injected electron beams and $L_B=-B/\nabla B$ is the scale height of the magnetic field. We take the average scale height $L_B$ of $\sim$10 Mm in the vicinity of the second harmonic gyroresonance layer above source I. Assuming $v_b=0.1$--0.5$c$ (typical for electron beams that produce type III radio bursts; see, \citealt{2014RAA....14..773R}), the expected frequency drift rate is $\left|\dot{\nu}\right|/\nu\approx3$--15 $\rm{s}^{-1}$. Therefore, the time delay for the ECM emission to traverse the 1--2 GHz band is expected to be $\Delta t=0.05$--0.23 s, which should have been detected by the VLA with a time resolution of 0.05 s, unless the electron beams have much larger speeds than those typically expected to drive the type III radio bursts (see, however, \citealt{1994A&A...286..611P}, \citealt{2003A&A...410..307K}, and \citealt{2018ApJ...866...62C} for a few cases in which $\gtrsim$0.5c electron beams are implied). For such a reason, we suggest that the repetitive injection of electron beams for the cause of the source I QPPs is deemed unlikely.

\begin{figure*}[!ht]
\plotone{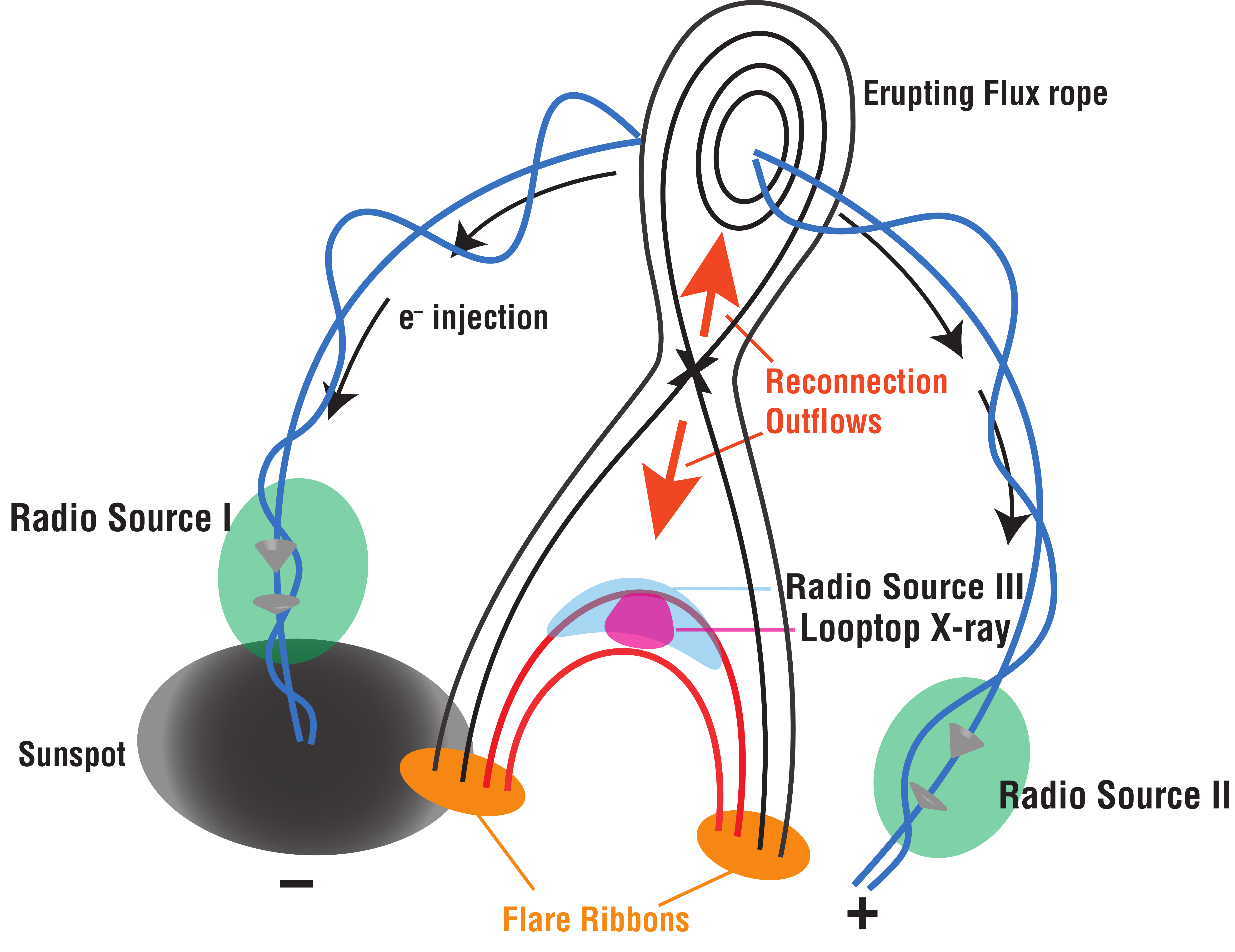}
\caption{Schematic diagram of the observations and a possible scenario of the physical processes that drives the observed QPP phenomena. An erupting magnetic flux rope led to magnetic reconnection in a current sheet as in the standard flare scenario. Nonthermal electrons that gain access to the erupting flux rope through reconnection can travel along the field line to the conjugate footpoints of the flux rope. These regions are unstable to ECM wave growth thanks to the presence of a loss-cone-type configuration and a favorable $\nu_{\rm{pe}}/\nu_{\rm{ce}}<1$ condition, producing the observed coherent radio sources I and II. Meanwhile, nonthermal electrons trapped in the looptop region produce X-ray bremsstrahlung and incoherent radio gyrosynchrotron emission, observed as a RHESSI looptop X-ray source and VLA radio source III, respectively. Modulation of the intensities of the spatially distinct sources may be due to different drivers, resulting in the observed QPPs with a variety of periodicities.\label{fig: cartoon}}
\end{figure*}

Different from the radio source I QPPs, radio source III and the looptop X-ray source share a longer period of $\sim$25--45 s. In Section \ref{sec: gs}, we have shown they coincide spatially and have a close time correlation, suggesting they are likely associated with the same trapped energetic electron population. Radio and X-ray QPPs with relatively longer periods are usually suggested to be associated with local MHD oscillations \citep{2006A&A...446.1151N,2012ApJ...748..140M}, periodic injections of nonthermal electrons \citep{2016AdSpR..57.1456K,2016SoPh..291.3427K,2021ApJ...910..123C}, or periodic energy release or electron acceleration \citep{1987SoPh..111..113A,2001ApJ...562L.103A}. \citet{2008ApJ...684.1433F} conducted a detailed investigation of the two models and compared their results using the observed radio and HXR emission properties including the spectral index and polarization. They concluded that the injection of nonthermal electrons, which leads to the variation of electron distribution instead of the magnetic field, was the more likely cause of the observed QPPs. On the other hand, the modulation of the looptop emission can also be owing to the modulation of local electron acceleration at the looptop region. \citet{2020ApJ...900...17Y} reported periodic impulsive looptop X-ray and microwave emission and demonstrated their close correlation with sun-ward plasma outflows. They suggest repetitive outflows can transport the released energy to accelerate the electrons in the looptop region. In addition, the looptop oscillations can be related to the ``magnetic tuning fork'' mechanism, driven by reconnection outflows impinging upon the loops below and creating an oscillating horn-shaped magnetic field structure \citep{2016ApJ...823..150T,2020ApJ...905..165R}. In our event, although complementary information from X-ray and radio imaging and spectroscopy (Figures \ref{fig: xray fit} and \ref{fig: gs fit}) has helped us identify the location and timing correlation as well as their underlying emission mechanisms, due to, in part, the pileup effect in the X-ray data, detailed spectral analysis of the nonthermal component is not possible. Also, as we discussed in Section \ref{sec: gs}, the spectral fitting for the radio looptop source is limited by the angular resolution and the frequency bandwidth of the radio instrument, which hampers us from obtaining more accurate constraints for the source parameters. Therefore, our data do not allow us to explicitly distinguish the different oscillation scenarios. 

Although the observed QPPs at the flux rope footpoint and the looptop can be due to a variety of mechanisms, an outstanding question is whether or not these QPPs can be explained by a common driver. We argue that it is unlikely the case because the periodicities of the observed QPPs differ by a factor of six to eight. Therefore, at least one, if not all, of the spatially distinct QPPs should result from oscillations in the local source rather than driven by a common external mechanism (e.g., periodic injections of electrons from the reconnection region).

To summarize, we propose a possible scenario summarizing the observations, illustrated in Figure \ref{fig: cartoon}. First, magnetic reconnection in a current sheet is induced by the eruption of a magnetic flux rope. Energetic electrons are accelerated at the reconnection site or, perhaps more likely, in the above-the-looptop region \citep[see, e.g.,][]{2020NatAs...4.1140C}. According to the scenario depicted by \citet{2020ApJ...895L..50C}, some upward-propagating nonthermal electrons can gain access to the erupting flux rope, possibly following the freshly reconnected magnetic field lines that join the flux rope, and then propagate downward toward the conjugate footpoints of the flux rope. The magnetic configuration near the flux rope footpoints features strongly converging magnetic field lines, and is favorable for developing a loss-cone-type distribution to generate coherent ECM emission through a nonlinear growth of $o$- and $x$-mode waves. The loss-cone distribution in the ECM-unstable region may be modulated by either wave-particle relaxation oscillation or the fast-mode sausage oscillation. The opacity above the magnetic polarities at the flux rope footpoints allows the $o$-mode to escape much more easily, which contributes, in part, to the strongly polarized $o$-mode emission at both locations. At the flare reconnection current sheet, nonthermal electrons are trapped at the top of the flare arcade. They generate incoherent gyrosynchrotron radio emission and X-ray bremsstrahlung emission, which are observed by VLA and RHESSI as the radio source III and X-ray looptop source, respectively. A quasiperiodic electron acceleration/injection or modulations of the looptop region can give rise to the observed QPP signatures of the looptop radio and X-ray source. However, because the periodicities of the looptop sources and the footpoint sources are vastly different, they cannot be explained by a single modulation mechanism. Instead, it must involve multiple mechanisms which operate in different magnetic loop systems and at different periods.

\acknowledgments

This work makes use of public VLA data from the observing program VLA/16A-377. The NRAO is a facility of the National Science Foundation (NSF) operated under a cooperative agreement by Associated Universities, Inc. This work constitutes part of Y.L.'s Ph.D. thesis supported by NSF CAREER grant AGS-1654382 awarded to B.C. of the New Jersey Institute of Technology. S.Y. is supported by NASA grants 80NSSC21K0623 to NJIT. R.S. acknowledges the support of the Swiss National Foundation, under grant
200021\_175832.

\facilities{EVLA, SDO, RHESSI, Fermi}

\software{CASA \citep{2007ASPC..376..127M},
          Astropy \citep{2018AJ....156..123A}, 
          SunPy \citep{2020ApJ...890...68S}.
          }

\bibliography{references}

\end{document}